\documentclass[prl,superscriptaddress,reprint,twocolumn,showpacs]{revtex4-1}

\usepackage{amsfonts} 
\usepackage{amsmath}
\usepackage{amssymb}
\usepackage{graphicx}
\usepackage{hyperref}
\usepackage{color}
\usepackage{natbib}

\newcommand{\angles}[1]{\left\langle{#1}\right\rangle}

\begin{document}

\title[Asymptotic Discord and Entanglement of Non-Resonant]{Asymptotic Discord and Entanglement of Non-Resonant Harmonic Oscillators in an Equilibrium Environment}

\author{Luis A. Correa}
\affiliation{IUdEA Instituto Universitario de Estudios Avanzados, Universidad de La Laguna, 38203 Spain}
\affiliation{Dpto. F\'{\i}sica Fundamental, Experimental, Electr\'{o}nica y
Sistemas, Universidad de La Laguna, 38203 Spain}

\author{Antonio A. Valido}
\affiliation{IUdEA Instituto Universitario de Estudios Avanzados, Universidad de La Laguna, 38203 Spain}
\affiliation{Dpto. F\'{\i}sica Fundamental II, Universidad de La Laguna, 38203 Spain}

\author{Daniel Alonso}
\email{dalonso@ull.es}
\affiliation{IUdEA Instituto Universitario de Estudios Avanzados, Universidad de La Laguna, 38203 Spain}
\affiliation{Dpto. F\'{\i}sica Fundamental, Experimental, Electr\'{o}nica y
Sistemas, Universidad de La Laguna, 38203 Spain}


\begin{abstract}
In this work, we calculate the exact asymptotic quantum correlations between two interacting non-resonant harmonic oscillators in a common Ohmic bath. We derive \emph{analytical formulas} for the covariances, fully describing any Gaussian stationary state of the system, and use them to study discord and entanglement in the strong and weak dissipation regimes. We discuss the rich structure of the discord of the stationary separable states arising in the strong dissipation regime. Also under strong dissipation, when the modes are not mechanically coupled, these may entangle only through their interaction with the \emph{common} environment. Interestingly enough, this stationary entanglement is only present within a \emph{finite band of frequencies} and increases with the dissipation rate. In addition, robust entanglement between \emph{detuned} oscillators is observed at low temperature.
\end{abstract}

\maketitle

Over the last decades, non-classical correlations such as \emph{discord} and \emph{entanglement} have been widely acknowledged to play a central role in quantum mechanics. In particular, the robustness of entanglement against decoherence in continuous-variable (CV) systems has been object of study in a number of recent publications \cite{plenio20021,PazPRL,*PazPRA,Galve,Distance,Ludwig,PhysRevLett.105.180501,CBM&IBM2007,IBMManiscalco,*PhysRevA.82.012313,PhysRevA.76.042127,refId0}. Due to the fact that entanglement is a valuable resource in quantum communication and information processing with CV, the investigation of the conditions under which bipartite states \emph{indefinitely} preserve their non-separable nature in presence of noise and dissipation is not only of fundamental importance, but also of practical interest. 

Gaussian two-mode states occupy a priviledged position among all entangled CV preparations, since they combine easy experimental realization (e.g. the output beams in an optical parametric oscillator), with simple mathematical description in terms of their second order moments, which are experimentally determined via homodyne 
detection \cite{PhysRevLett.102.020502}. These CV systems also find physical realization in mechanical oscillators, such as trapped ions coupled by their electrostatic interaction \cite{brown20111,*harlander2011trapped}. A simple but rather general model describing those systems consists of two coupled detuned harmonic oscillators, linearly interacting, and in contact with a bath.

The stationary entanglement of two resonant modes in a squeezed state under decoherence was studied in Refs. \cite{PazPRL,*PazPRA}, whereas dynamical features of the non-resonant case were addressed in \cite{Galve}. Additionally, different \emph{long-time} behaviours were experimentally observed in \cite{barbosa2010robustness} after simulating decoherence on one of the modes. Other recent publications such as \cite{CBM&IBM2007,IBMManiscalco,*PhysRevA.82.012313} treat the dynamics of quantum correlations in similar systems for a variety of structured environments. Nonetheless, it must be noted that the majority of these works rely on \emph{weak} system-environment coupling assumption. 

On reference to quantum discord \cite{zurek20001, *henderson20011, *olliver20011}, it was introduced to capture the \emph{quantumness} of correlations. Since the majority of quantum states, including most of the separable ones, have non-zero discord \cite{ferraro20101}, it is said to be more general than entanglement as a measure of non-classicality. Discord has recently attracted attention in the field of quantum information, now that for some computational models wich make no use of entanglement, it is believed to provide exponential speedup over their best known classical counterparts \cite{PhysRevLett.100.050502,*PhysRevLett.101.200501}.

In this Letter, we provide \emph{exact analytical formulas} fully characterizing the Gaussian stationary state of two interacting non-resonant modes in contact with a common thermal environment by solving the quantum Langevin equations in the asymptotic limit \cite{Distance, Ludwig}. We then compute quantum discord and entanglement exactly, being our main interest the robustness of stationary quantum correlations against temperature and dissipation, in the less studied case of finitely \emph{detuned} modes.

Our system consists of two harmonic oscillators coupled
through a term bilinear in the coordinates $H_{S}=\sum\nolimits_{i=1}^{2}\frac{p_{i}^{2}}{2}+\frac{1}{2}\omega_{i}^{2}x_{i}^{2}+\frac{k}{2}\left(  x_{1}-x_{2}\right)  ^{2}$. The interaction with a thermal environment 
can be modeled by coupling
them to a common bosonic heat bath, which yields: 
$H=H_{S}+\sum_{\mu}\left(  \frac{p_{\mu}^{2}}{2m_{\mu}}+\frac{1}{2}m_{\mu}%
\omega_{\mu}^2x_{\mu}^{2}\right) +\mathcal{F}\left(  x_{1},x_{2}\right)
\sum_{\mu}g_{\mu}x_{\mu}+\mathcal{F}\left(  x_{1},x_{2}\right)  ^{2}\Omega
^{2}/2$ \footnote{The last term is introduced to compensate the
renormalization appearing as a result of the system-bath coupling.}, where
$\Omega^{2}$ is given by $2\int_{0}^{\infty}d\omega~J\left(  \omega\right)
/\omega$, and $J\left(  \omega\right)$ is the bath spectral density \cite{weiss2008quantum}. We use units in which $\hbar=k_B=1$.

As noted in Refs. \cite{PazPRL,PazPRA,CBM&IBM2007} among others, under
suitable conditions, an initially separable state may become entangled in this model, 
even if $k=0$. However, this ability of the \emph{common} environment to create non-classical correlations
between the two parties highly depends on the physical distance $r$ that
separates them \cite{Distance,refId0}, so that \emph{distant} dissipative oscillators
would be better modeled by considering \emph{independent} identical environments.

In the following, all averages are taken with respect to a separable initial state
in the system's and bath's degrees of freedom $\rho\left(  0\right)
\otimes\rho_{T}$ where $\rho\left(  0\right)  $ is a Gaussian state of the
two oscillators, and $\rho_{T}$ is a thermal equilibrium state of the
environment at temperature $T$.

By choosing $\mathcal{F}\left(  x_{1},x_{2}\right)  =x_{1}+x_{2}$ in $H$ 
one may derive the following quantum Langevin equation \cite{weiss2008quantum}:%
\vspace{-0.25cm}
\begin{eqnarray}
\label{lagevin1}
\ddot{x}_{i}&+&\left(  \omega_{i}^{2}+\Omega^{2}\right)  ~x_{i}+\sum_{j=1}%
^{2}\left[  k\left(  x_{i}-x_{j}\right)  +\Omega^{2}x_{j}\right]
\tilde{\delta}_{ij}
\\ \nonumber &=&F\left(  t\right)  +\sum_{i=1}^{2}\int_{-\infty}%
^{t}ds~\chi\left(  t-s\right)  ~x_{i}\left(  s\right)  \text{,} \,\, 
\end{eqnarray}
where $\tilde{\delta}_{ij}\equiv1-\delta_{ij}$.
The quantum force $F\left(t\right)  $ acting on both oscillators and the
dissipative kernel $\chi\left(  t\right)  $ are connected by the Kubo relation
$\chi\left(  t\right)  =i\theta\left(  t\right)  \left\langle \left[  F\left(
t\right)  ,F\left(  0\right)  \right]  \right\rangle $, where $\theta\left(
t\right)  $ stands for the Heaviside step function. For this choice of $\mathcal{F}\left(  x_{1},x_{2}\right)  $, the overall
system Hamiltonian $H$ is quadratic in positions and momenta, 
so that for $\omega_{1}\neq\omega_{2}$ the asymptotic state (which is
independent of $\rho\left(  0\right)  $) is guaranteed to be Gaussian. In the
case of resonant oscillators, the normal mode decomposition leads to an
effective decoupling of the relative degree of freedom from the bath, and thus
the stationary state would depend on the initial condition \cite{PazPRA}. In
the following we shall restrict ourselves to non-resonant oscillators.

We can solve Eq. (\ref{lagevin1}) by Fourier transforming and denoting 
$\hat{f}\left(  \omega\right)  \equiv\int_{-\infty}^{\infty
}dt~e^{i\omega t}f\left(  t\right)  $. Due to the linearity of the problem, this formally yields: 
$\hat{x}_{i}\left(  \omega\right)  =\sum_{j}\alpha_{ij}\left(  \omega\right)
\hat{F}\left(  \omega\right)  \text{,}$
where the generalized susceptibilities $\alpha_{ij}\left(  \omega\right)  $
are elements of the inverse of the matrix: 
$(\alpha^{-1})_{ii}=\omega_{i}^{2}+\Omega^{2}-\omega^{2}+k-\hat{\chi}(\omega)$ and
$(\alpha^{-1})_{ij}=-k+\Omega^{2}-\hat{\chi}(\omega)$ for $i \neq j$.

The central object of our study is the asymptotic covariance matrix
$\mathbf{\Gamma}$, defined as $\Gamma_{ij}\equiv\left\langle R_{i}%
R_{j}\right\rangle -\left[  R_{i},R_{j}\right]  /2$, with $\mathbf{R}%
\equiv\left\{  x_{1},p_{1},x_{2},p_{2}\right\}  $ which, up to local displacements, 
fully characterizes any Gaussian state of our bipartite system.
Provided with the susceptibility matrix $\mathbf{\alpha}\left(  \omega\right)
$, we may compute the power spectrum $\left\langle x_{i}x_{j}\right\rangle
_{\omega}\equiv\int_{-\infty}^{\infty}dt~e^{i\omega\left(  t-t^{\prime}\right)
}\left\langle x_{i}\left(  t-t^{\prime}\right)  x_{j}\left(  0\right)
\right\rangle $ of the stationary two-time correlation $\left\langle x_{i}
\left(  t-t^{\prime}\right)x_{j}\left(  0\right)\right\rangle=\left\langle
x_{i}\left(  t\right)  x_{j}\left(  t^{\prime}\right)  \right\rangle$ in terms of the power 
spectrum of the environmental force:
\vspace{-0.45cm}
\begin{equation}
\left\langle x_{i}x_{j}\right\rangle _{\omega}=\sum\nolimits_{k,l}\alpha_{ik}\left(
\omega\right)  \alpha_{jl}\left(  -\omega\right)  \left\langle FF\right\rangle
_{\omega}\text{.} \label{xixj}%
\end{equation}
The rest of power spectra $\left\langle R_{i}R_{j}\right\rangle _{\omega}$,
can be readily obtained from Eq. (\ref{xixj}) since
$\left\langle p_{i}x_{j}\right\rangle _{\omega}=-i\omega\left\langle
x_{i}x_{j}\right\rangle _{\omega}$ and $\left\langle p_{i}p_{j}\right\rangle
_{\omega}=\omega^{2}\left\langle x_{i}x_{j}\right\rangle _{\omega}$. The full
asymptotic covariance matrix then reads $\Gamma_{ij}\left(t,t\right)  
=\int_{-\infty}^{\infty} (d\omega/2 \pi)~\left\langle R_{i}R_{j}
\right\rangle _{\omega}-\left[R_{i},R_{j}\right]/2.$

To proceed further in deriving analytical expressions, 
we shall choose an Ohmic spectral density with
Lorentz-Drude cutoff $J\left(  \omega\right)  =2\gamma \omega /\pi(1+\omega^{2}%
/\omega_{c}^{2})$, that leads to a dissipative kernel $\hat{\chi}\left(  \omega\right)
=2\gamma\omega_{c}^{2}/(\omega_{c}-i\omega)$, and a bath correlation function 
$\left\langle FF\right\rangle _{\omega}=\pi J\left(
\omega\right)  \left[  1+\coth\left(  \omega/2T\right)  \right]$, where $\omega_c$ is the cutoff frequency and $\gamma$ the dissipation rate. The renormalization 
constant appearing in Eq. (\ref{lagevin1}) is $\Omega^2=2\gamma\omega_{c}$ 
for this choice of $J\left(\omega\right)$.
We start from Eq. (\ref{xixj}) to get
\vspace{-0.25cm}
\begin{equation}
\left\langle x_{i}\left(  t\right)  x_{j}\left(  t\right)  \right\rangle
=-\int_{-\infty}^{\infty}\frac{d\omega}{2\pi}\frac{g_{ij}\left(
\omega\right)  \coth\left(  \frac{\omega}{2T}\right)}{h\left(  \omega\right)  h\left(  -\omega\right)  }
  \text{,}
\end{equation}
where $g_{ij}\left(  \omega\right)  $ and $h\left(  \omega\right)  $ are 5th
order polynomials depending on $\omega_1,\omega_2,\omega_c,\gamma$ and $k$. 
After a lengthly calculation, these integrals can be solved to yield \footnote{See Supplemental Material at [URL will be inserted by publisher] for details.}:
\vspace{-0.25cm}
\begin{eqnarray*}
	\left\langle x_{i}\left(  t\right)  x_{j}\left(  t\right)  \right\rangle &=& \angles{x_i x_j}_{T\rightarrow\infty}-\sum\limits_{k=1}^5 a_k^{\left(i,j\right)}\psi\left(1-\frac{iz_k}{2\pi T}\right) \\
\left\langle p_i\left(t\right) p_j\left(t\right) \right\rangle &=& \angles{p_i p_j}_{T\rightarrow\infty}-\sum\limits_{k=1}^5 b_k^{\left(i,j\right)} \psi\left(1-i\frac{z_k}{2\pi T}\right)\text{,}
\end{eqnarray*}
while the remaining convariances are all zero. $a_{k}^{(i,j)}$ and $b_k^ {(i,j)}$ are temperature-independent coefficients for which we have explicit formulas, $z_k$ are the five complex roots of $h\left(\omega\right)$, and 
$\psi$ stands for the logarithmic derivative of Euler's gamma function. $\angles{R_i R_j}_{T\rightarrow\infty}$ denotes \emph{classical} mean values, to which the summations are \emph{quantum corrections}. 

To gain some insight into the physics of the problem, we will set $k=0$ and place ourselves in the limit of $\gamma/\omega_{c}\ll1$. To first order in this small parameter, the roots $z_{k}$ become 
$\pm\omega_{i}\pm \frac{\gamma \omega_{c}}{\left(\omega_{i}\mp i\omega_{c}\right)}$ and 
$i\omega_{c} -i2\gamma\omega_{c}^{2}\left(  \omega_{1}^{2}+\omega_{2}^{2}+2\omega_{c}
^{2}\right)/\prod_i (\omega_{i}^{2}+\omega_{c}^{2})$.
Additionally, in the very low
temperature regime (i.e. $T\ll\omega_{i}$) the asymptotic expansion
$\psi\left(  1+z\right)  \simeq\log z+\left(  2z\right)  ^{-1}$ may be
justified. Inserting all this into the formulas of the $\Gamma_{ij}$, and 
retaining terms only to first order in $T/\omega_{c}$,
$\omega_{i}/\omega_{c}$ and $\gamma/\omega_{c}$ one gets the
temperature-independent expressions:%
\vspace{-0.15cm}
\begin{eqnarray}
\left\langle x_{i}^{2}\right\rangle  &&  \simeq\frac{1}{2\omega_{i}}%
-\frac{\gamma}{2\pi}\left[  \frac{2\omega_{c}-\pi\omega_{i}}{\omega_{i}%
^{2}\omega_{c}}+\frac{4}{\omega_{c}^{2}}\left(  \log\frac{\omega_{c}}%
{\omega_{i}}\right)- \frac{1}{2}  \right]  \nonumber\\
\left\langle p_{i}^{2}\right\rangle  &&  \simeq\frac{1}{2}\omega_{i}%
+\frac{\gamma}{2\pi}\left[  3\pi\frac{\omega_{i}}{\omega_{c}}-2+4\log
\frac{\omega_{c}}{\omega_{i}}\right] \nonumber \\
\left\langle x_{1}x_{2}\right\rangle  &&  \simeq\frac{\gamma}{\pi\left(
\omega_{1}^{2}-\omega_{2}^{2}\right)  }2\log\frac{\omega_{2}}{\omega_{1}%
}\nonumber \\
\left\langle p_{1}p_{2}\right\rangle  &&  \simeq\frac{\gamma\log16}{2\pi}%
+\frac{\gamma \left(
\omega_{1}^{2}\log\frac{\omega_{c}^{2}}{4\omega_{1}^{2}}-\omega_{2}^{2}%
\log\frac{\omega_{c}^{2}}{4\omega_{2}^{2}}\right)}
{\pi\left(\omega_{1}^{2}-\omega_{2}^{2}\right)}  \text{.}  \nonumber
\end{eqnarray}
By setting $\gamma=0$, the variances $\left\langle x_{i}^{2}\right\rangle $
and $\left\langle p_{i}^{2}\right\rangle $ become those of the ground state of
the corresponding oscillators without dissipation, and the asymptotic
(separable) state has a diagonal covariance matrix. Taking into account that
$\omega_{c}/\omega_{i}\gg1$, one sees that increasing $\gamma$ leads to a
reduction of $\left\langle x_{i}^{2}\right\rangle $ while $\left\langle
p_{i}^{2}\right\rangle $ gets larger, as expected.
The terms proportional to $\log\omega_{c}/\omega_{i}$ in $\left\langle p_{i}p_{j}\right\rangle$, 
also occur in the same region of parameters for a single damped
oscillator \cite{weiss2008quantum}.

\begin{figure}[t]
\includegraphics[width=4.45cm]{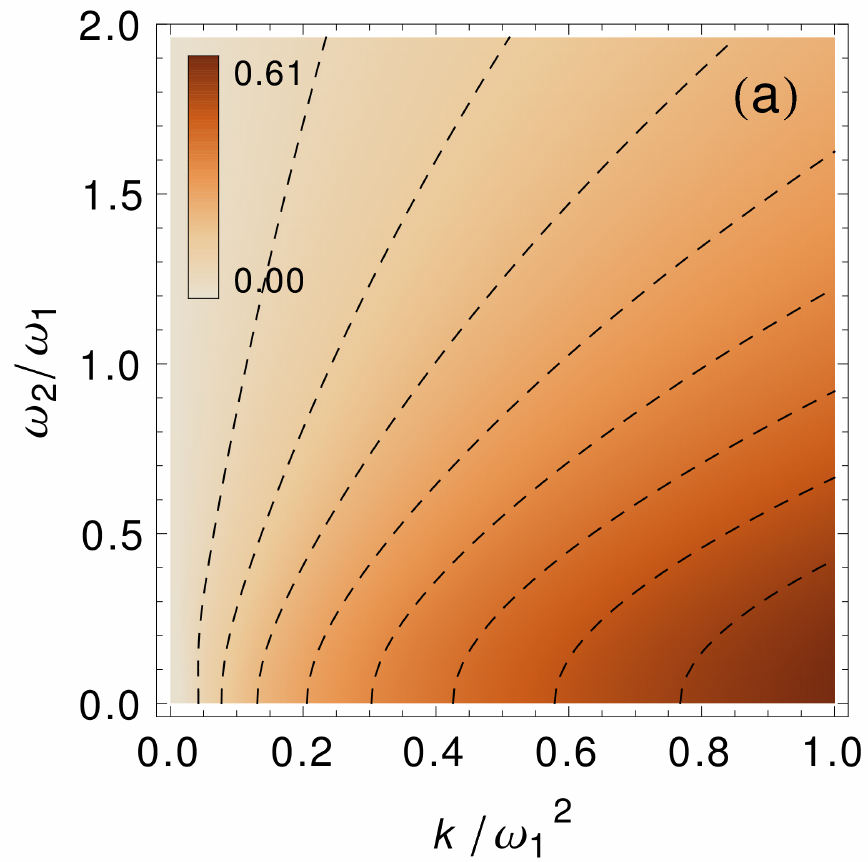} 
\includegraphics[width=4.1cm]{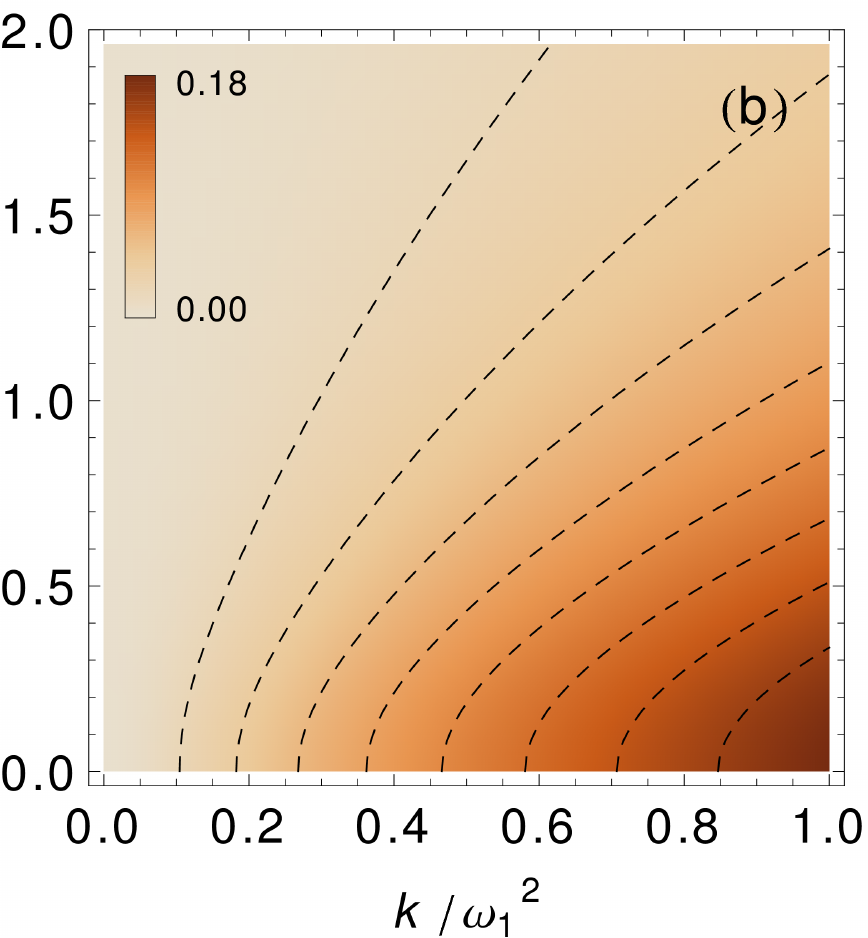}

\caption{(a) Asymptotic entanglement and (b) discord in the weak coupling regime
as functions of $k$ and $\omega_{2}$. We have chosen
$\omega_{1}=10$, $\gamma=0.01$, $\omega_{c}=500$ and $T=0.5$. Both measures of quantum correlations exhibit a similar
behavior: They increase with $\omega_{2}/k\rightarrow0$. Dashed lines are contours and the color legends are plotted in the upper left corners.\label{figure1}}
\end{figure}

These analytical and \emph{exact} expressions for the asymptotic covariances $\Gamma_{ij}$
are one of the main results of this Letter. From them, we shall compute all quantum correlations without \emph{any assumptions} on the different time scales involved in the problem. The \emph{logarithmic negativity} \cite{LogarithmicNegativity} is a suitable measure of entanglement is this case. It is based on the \emph{positivity of the partial transpose} separability criterion \cite{peres1996separability,*horodecki1996separability}, which is necessary and sufficient for bipartite Gaussian states \cite{Simon}. When it comes to discord, it is esentially the difference between the mutual information of the bipartite state, before and after performing a complete measurement on one of the parts (in what follows, mode two). Even though its calculation is usually very involved, if one restricts to Gaussian generalized measurements it is straightforward to compute for bipartite Gaussian states, as recently shown in \cite{giorda2010gaussian,*adesso2010quantum, PhysRevA.82.012313,isar2011quantum}. We now turn to present the results thus obtained:

\paragraph{Weak dissipation.---\label{weak coupling}}
We will look first into the weak dissipation regime, where the relaxation time scale
$\tau_{R}=\gamma^{-1}$ is much larger than the bath correlation time, 
given by $\tau_{B}=\nu_{1}^{-1}$ whenever $\omega_{c}>\nu_{1}$ \cite{breuer2002theory}. We shall
also pick a low temperature $T$ for which non-zero asymptotic entanglement might exist. Results 
are presented in Fig. \ref{figure1}. In accordance with \cite{PazPRA}, we observe $E_{\mathcal{N}}\neq 0$ \emph{out of resonance} at low temperatures, and overall similar behaviour of entanglement and discord.

In the limit $k\gg\omega_{i}$, the oscillators, whose effective frequency $\tilde{\omega}^2$ is given by 
$\omega_{i}^2+k+2\gamma\omega_c$, become nearly resonant and thus, approximately decouple in the variables
$\eta_{\pm}\equiv\left(x_{1}\pm x_{2}\right)/\sqrt{2}$ and $\pi_{\pm}\equiv\left(p_{1}\pm p_{2}\right)/\sqrt{2}$. As $\omega_{i}/k\rightarrow0$,
$\left\lbrace\left\langle\eta_{+}^2\right\rangle,
\left\langle\pi_{-}^2\right\rangle\right\rbrace\rightarrow\infty$ while
$\left\lbrace\left\langle\pi_{+}^2\right\rangle,
\left\langle\eta_{-}^2\right\rangle\right\rbrace\rightarrow0$ \cite{Ludwig}. This corresponds
to a non-symmetric two-mode squeezed thermal state with infinite squeezing $r$, i.e. the ideal
EPR (maximally entangled) state \cite{einstein1935can}. 

\begin{figure}[t]
\includegraphics[width=4.16cm]{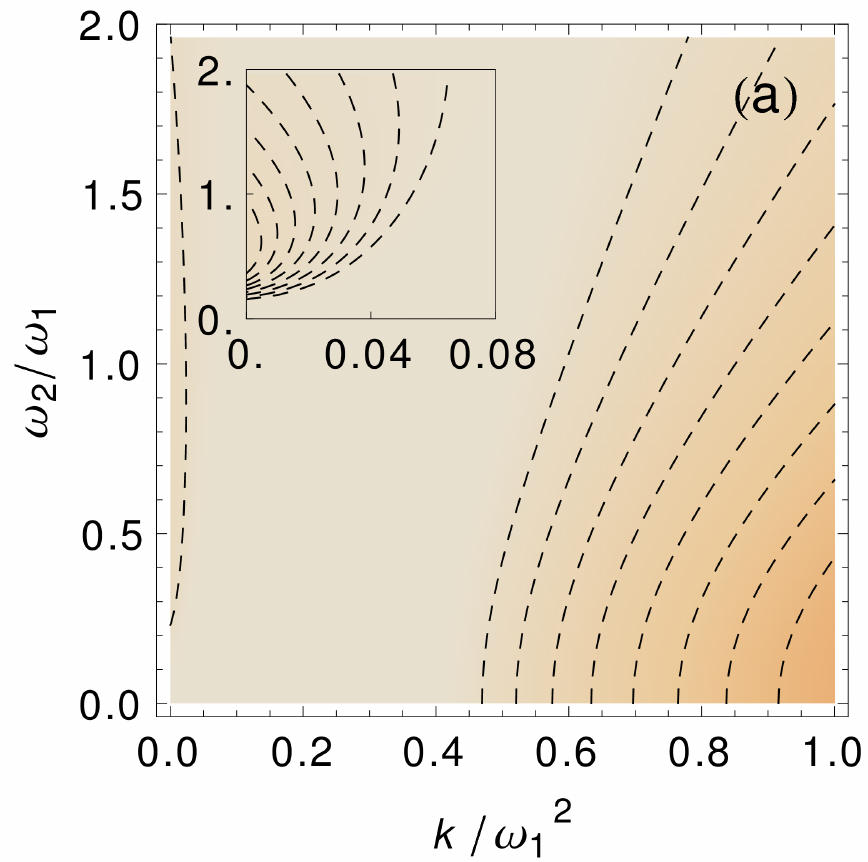} 
\includegraphics[width=4.39cm]{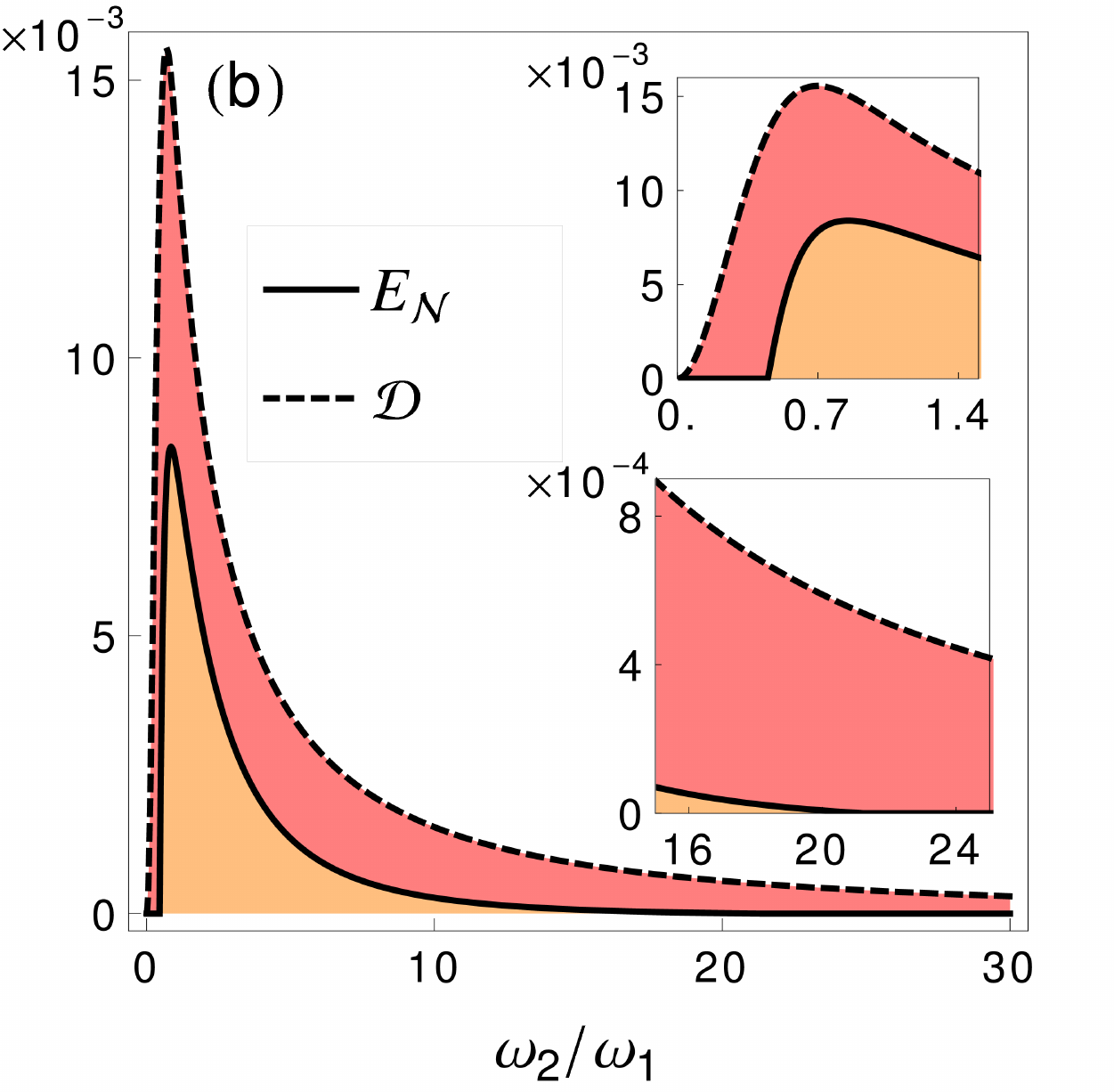} 

\caption{\protect\label{figure2}(a) Asymptotic entanglement in the strong dissipation regime as a function of $k$ 
and $\omega_{2}$. Due to the bigger decoherence entanglement is mostly lost, though its maximum is still attained as 
$\omega_{2}/k\rightarrow0$. Interestingly, in the inset of (a), we can observe a small region around $k=0$ where non-classical correlations are \emph{enhanced} by the dissipation. This purely
non-Markovian effect, enters into our equations through the 
renormalization term $\Omega^2 x_{1} x_{2}$ in the Hamiltonian $H$, which for sufficiently high dissipation rate, results in an effective environment-assisted coupling between the oscillators.
(b) Asymptotic entanglement (continuous line) and discord (dashed line) 
as functions of $\omega_{2}$ 
alone for $k=0$. For both plots, parameters are the same as in 
Fig. \ref{figure1} except for the dissipation $\gamma=0.5$.}
\end{figure}

\paragraph{Strong dissipation.---\label{strong coupling}}
We achieve this regime by increasing $\gamma$, so that the relaxation time $\tau_{R}$
becomes comparable to the bath correlation time $\tau_{B}$. Results are presented and summarized in 
Fig. \ref{figure2}.  

Even at $k=0$, there may exist non-vanishing stationary entanglement. The
corresponding asymptotic states are thus non-Gibbsian, as could be expected at sufficiently low 
temperatures \cite{nieuwenhuizen20021}. Actually, one sees in Fig. \ref{figure2} (b) that for fixed $\omega_{1}$, $\gamma$, $T$ and $\omega_{c}$, 
there exists a \emph{band} of frequencies $\omega_{2}\in\left[\omega',
\omega''\right]$ for which the asymptotic state is non-separable. As the temperature increases,
the amount of entanglement gets reduced, while $\omega'$ and $\omega''$ come closer to each other. On the
contrary, by increasing the dissipation rate $\gamma$, the maximum attainable $E_{\mathcal{N}}$ 
and the bandwidth $\omega''-\omega'$ get larger. Reducing the 
frequency $\omega_{1}$ also leads to narrower bandwidth.
One may also see that $\omega'$ is almost insensitive to changes in $\omega_{c}$, 
while $\omega''$ slightly decreases as $\omega_{c}$ grows.

The simple model $H_{S}=\sum\nolimits_{i}[p_{i}^2/2+(\omega_{i}^2+\Omega^2) x_{i}^2/2]+
\Omega^2x_{1} x_{2}$ proves useful to understand these features. Note that 
the renormalization $\Omega^2=2\gamma\omega_{c}$ has been introduced to account for the bath-mediated 
coupling. If $w_{i}/\Omega^2\ll1$, the oscillators are again near resonance
and approximately decouple in the coordinates $\{\eta_{\pm},\pi_{\pm}\}$. 
Then, $E_{\mathcal{N}}\neq0$ whenever
$\tilde{\nu}_{-}^2=\left\langle\eta_{-}^2\right\rangle \left\langle\pi_{+}^2\right\rangle<1/4$ \cite{Ludwig}.

It is easy to see that the lowest symplectic eigenvalue of the partially transposed covariance matrix $\tilde{\nu}_{-}$ grows
with $T$ and decreases with $\Omega^2$ and $\omega_{i}$. Thus, an increase in the temperature must be 
compensated by an increase in $\omega'$ for the inequality to be satisfied. Conversely,
an increase in either $\Omega^2$ or $\omega_{1}$ allows for a reduction of the critical frequency $\omega'$. The existence of 
an upper bound $\omega''$ for this non-separable region
is related to the structure of the bath, so that its behavior can not be inferred from this simple 
non-dissipative model. 

Finally, we also remark that in our symplified model, $\tilde{\nu}_{-}$  
decreases as $|\Omega^2-k|$ grows, since the effective coupling between the oscillators is the net result of two competing effects, that may eventually cancel each other: The environment-assisted coupling and the direct interaction between the oscillators. 
This explains qualitatively why $E_{\mathcal{N}}=0$ for \emph{intermediate} couplings in 
Fig. \ref{figure2}.
 
\begin{figure}[t]
\includegraphics[width=5cm]{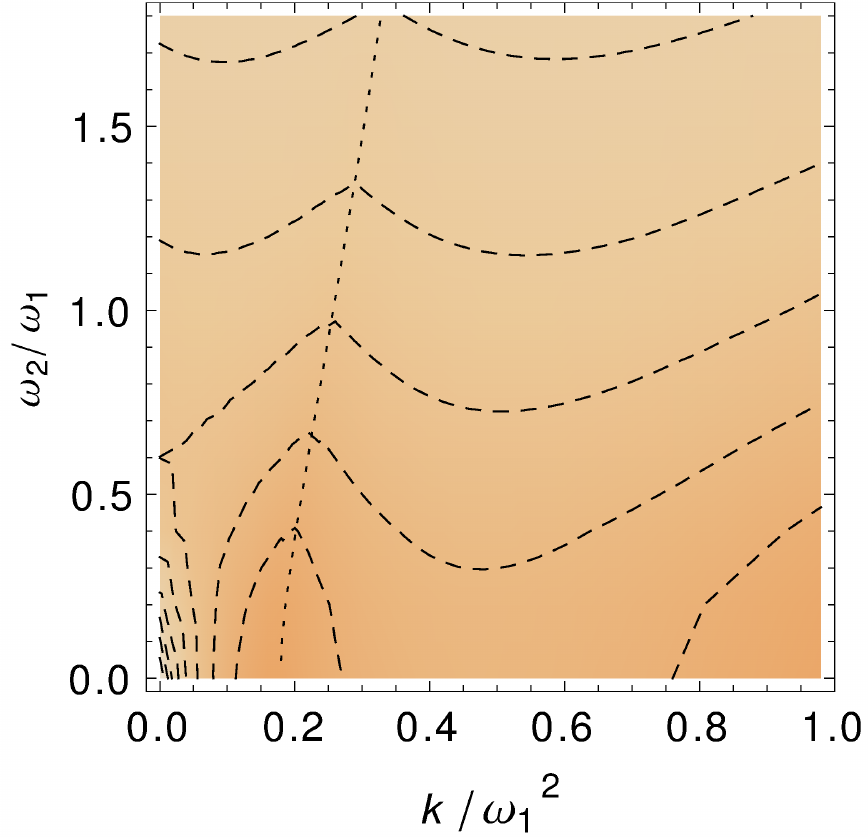}

\caption{The asymptotic discord displays a very different behaviour in 
the strong dissipation regime. Since the entanglement has considerably decreased, it no longer
dominates over other quantum correlations and, as could be expected, remains non-zero even 
when the asymptotic state is separable. It displays a
maximum around $k\simeq20$, marked with the dotted line, and also vanishes in the limit  $\left\lbrace k,\omega_2\right\rbrace\rightarrow 0$ (see text for discussion).
All parameters are the same as in Fig. \ref{figure2}.
\label{figure2c}}
\end{figure}

Finally, Fig. \ref{figure2c} contains results on discord in this regime. The global purity $\mu$
may be found to be roughly constant for all stationary states, except those with $\left\lbrace k,\omega_2\right\rbrace\rightarrow 0$, while $\tilde{\nu}_{-}$, is maximum on the dotted line of figure \ref{figure2c}. This suggest that at fixed $\mu$, Gaussian quantum discord increases with 
$\tilde{\nu}_{-}$, which is consistent with \cite{giorda2010gaussian}, where this was seen to hold for squeezed thermal states. Additionally, since the partial purity $\mu_2\rightarrow 0$ as $\left\lbrace k,\omega_2\right\rbrace\rightarrow 0$ the marginal stationary states $\rho_2 \equiv\hbox{Tr}_1(\rho_{12})$ become maximally mixed in this limit, therefore yielding a quantum-classical state \cite{PhysRevLett.100.090502} with zero discord as revealed by measurements on mode two.  

At higher temperatures and stronger dissipation rates, the fate of the asymptotic entanglement is to
disappear completely, in contrast with the asymptotic quantum discord, which is always non-zero as could be expected \cite{ferraro20101}.

To summarize and conclude, we have exactly solved the problem of the
asymptotic quantum correlations in a system of two coupled non-resonant
harmonic oscillators in a common bath. We provided exact
analytical formulas for the covariances, fully characterizing any Gaussian
stationary state of the system, and used them to compute the asymptotic 
entanglement and discord, in the weak and strong dissipation regimes.
We found that stationary entanglement between non-resonant oscillators may exist at sufficiently low
temperatures. Furthermore, for a finite band of frequencies, it may be created only due to the environment-assited interaction between the two modes in the strong dissipation regime. Also in this regime we discuss the non trivial structure that quantum discord shows when the stationary states are separable.

The authors are grateful to Jos\'{e} P. Palao for fruitful discussion.
Financial support from the Spanish MICINN (FIS2010-19998) and from the ACIISI fellowships (85\% cofunded by European Social Fund) is also gratefully acknowledged. 
D.A. is grateful to E. Sent\'{\i}s Rodr\'{\i}guez for encouragement.

%

\end{document}